\documentclass[%
 reprint,
nofootinbib,
 amsmath,amssymb,
 aps,
]{revtex4-2}

\usepackage{graphicx}
\usepackage{dcolumn}
\usepackage{bm}
\usepackage[colorlinks=true, linkcolor=blue, citecolor=blue, urlcolor=blue]{hyperref}

\UseRawInputEncoding

\usepackage{lineno}
\usepackage{amssymb}
\usepackage{amsmath}
\usepackage{braket}
\usepackage{adjustbox}
\usepackage{array}
\usepackage{xcolor}

\newcommand{\sub}{\rm{sub}}
\newcommand{\rmmin}{\rm{min}}
\newcommand{\rmmax}{\rm{max}}
\newcommand{\obs}{\rm{obs}}
\newcommand{\NFW}{\rm{NFW}}

\newcommand{\ddedect}{d_{\rm{detect}}}

\begin{document}

\preprint{APS/123-QED}

\title{Darkness in the Crust: Searching for “truly” Dark Subhalos with Paleo-detectors}

\author{Xiuyuan Zhang}
 \email{xiuyuan@mit.edu}
\author{Lina Necib}%
\affiliation{%
Physics Department and Kavli Institute for Astrophysics and Space Research, Massachusetts Institute of Technology, Cambridge, MA 02139, USA
}%

\author{Denis Erkal}
\affiliation{
School of Mathematics and Physics, University of Surrey, Guildford GU2 7XH, UK}%

\date{\today}

\begin{abstract}
Low-mass dark matter (DM) subhalos are pivotal in understanding the small-scale structure of the universe, thereby offering a sensitive method to discriminate between different cosmological models. 
In this study, we estimate the local number density of cold DM subhalos in the solar neighborhood, and demonstrate that their sparse distribution makes their detection via direct detection experiments highly improbable. 
However, it is plausible to expect that an $\mathcal{O}(1)$ number of subhalos could be detected by Paleo-detectors, a proposed new technique to look for DM by reading out damage tracks left by past DM interactions in minerals, due to their extended exposure times. 
Hence, we explore how Paleo-detectors can serve as effective probes for the properties of low-mass subhalos, $\mathcal{O}(10^{-5}-10^8) M_{\odot}$. 
We find that Paleo-detectors might be able to constrain certain regions of the subhalo mass-concentration relation (for subhalo masses of $10-10^4 M_\odot$ if DM has a mass of $\sim5$GeV). 
This is a new and complementary type of study that seeks to combine information from the particle nature of DM to that of small scale structures. 
\end{abstract}

\maketitle

\section{\label{sec:level1}Introduction}

Lambda cold dark matter ($\Lambda$CDM) is widely considered as the standard model of Big Bang cosmology \cite{PhysRevD.23.347, STAROBINSKY1982175, HAWKING1982295, PhysRevD.28.679}, primarily because it effectively explains key phenomena such as the peaks of the cosmic microwave background (CMB) \cite{Hinshaw_2009, 2016}, the distribution of large scale structures \cite{Wu_1998, Refregier_2003}, and the accelerating expansion of the universe \cite{Abbott_2019, 2020}. Central to $\Lambda$CDM is the concept of dark matter (DM), an elusive substance with no known interactions with ordinary matter other than its gravitational effects. The existence and particle nature of DM are therefore crucial in testing the cosmological models that account for the structural formation of the universe.

DM can be detected in multiple ways, including DM production at colliders~\cite{ hooper2018tasilecturesindirectsearches}, indirect detection experiments~\cite{Steigman_2012, slatyer2017tasilecturesindirectdetection, hooper2018tasilecturesindirectsearches}, and direct detection experiments~\cite{PhysRevD.30.2295, PhysRevD.31.3059, PhysRevD.33.3495, PhysRevD.37.1353, COLLAR1992181}. Direct detection is a type of terrestrial experiment that measures the recoil of nuclei from scattering with DM particles, primarily focusing on a popular candidate called Weakly Interacting Massive Particles (WIMPs). WIMPs are hypothetical particles that interact with standard model particles with a strength as weak as or weaker than the weak force scale and have a mass at the GeV to TeV scale~\cite{JUNGMAN1996195, Garrett_2011}. Various direct detection experiments like XENON~\cite{Aprile_2016, PhysRevLett.121.111302, Aprile_2019}, CRESST~\cite{ Angloher_2016, cresstcollaboration2017resultslowmassdarkmatter}, DarkSide~\cite{Agnes_2018, Agnes_2018_01}, DARWIN~\cite{Aalbers_2016}, CYGNUS~\cite{vahsen2020cygnusfeasibilitynuclearrecoil}, SuperCDMS~\cite{Agnese_2017}, etc. have set tight constraints on the parameter space of spin dependent and independent interactions with nucleons for GeV-mass DM particles, with LZ~\cite{mount2017luxzeplinlztechnicaldesign, aalbers2024darkmattersearchresults} setting the current world-leading bounds, pushing the sensitivity regions to be within reach of the neutrino floor. With the proposed collaboration between XENON, LZ and DARWIN (XLZD), the joint effort is anticipated to advance the constraints even further, and ultimately reach the neutrino floor. 
If DM continues to evade direct detection, it will become critical to develop more innovative strategies to search for DM and test $\Lambda$CDM. 

One of the distinctive predictions of $\Lambda$CDM is the abundant presence of small-scale substructures within a host halo, referred to as DM subhalos~\cite{10.1093/mnras/183.3.341, 2008Natur.454..735D, 2008MNRAS.391.1685S, Klypin_2011}. These subhalos are thought to have integrated into the main halo through the process of hierarchical merging; halos grow through successive mergers in a process called hierarchical structure formation \citep{Klypin_1999, Diemand_2007, 2008MNRAS.391.1685S}. The more massive halos host galaxies, and their mass can therefore be inferred from the mass of their observed baryonic components through a so-called galaxy-halo connection \cite{Conroy_2006,Wechsler_2018}. Smaller halos, however, are not massive enough to host galaxies. Recent simulations suggest that the threshold mass for galaxy formation might be around $10^8M_{\odot}$ ~\cite{Nadler_2024}. However, confirming this finding through observational data presents significant challenges: Galaxies at total of masses $\lesssim 10^9 M_{\odot}$ are referred to as ultrafaint dwarf galaxies, and are extremely dim and hard to observe~\cite{Hezaveh_2016, Nadler_2021, _eng_l_2023, tsang2024substructuredetectionrealisticstrong}. Additionally, there are substantial theoretical uncertainties at the lower mass limits of the galaxy-halo connection \cite[see e.g.][]{Wechsler_2018}, further complicating the validation of these simulation results. 

Whatever the cutoff might be, the smallest of halos do not host any visible matter, and are hence truly ``invisible" to us. At the same time, these dark subhalos provide crucial insights into the nature of DM and its distribution, specifically through their tight connection to the initial power spectrum \cite{Zentner_2003}. The subhalo masses and density profiles can hint at the mass-concentration relation $c(M)$ \cite{Correa_2015}, which describes how the density of a halo increases towards its center as a function of its total mass. Additionally, their abundance helps test and refine the lower tail of the subhalo mass function \cite{2008MNRAS.391.1685S}. Both aspects are essential for distinguishing between different DM models. 


One method to probe these dark subhalos is through strong lensing of a galaxy-galaxy system~\cite{Nadler_2019, Hsueh_2019, Nadler_2021, Mao_2021, Das_2021, Maamari_2021, Nadler_2021_01, Wagner_Carena_2023}. One of the galaxies serves as a ``main deflector" to deflect the light emitted by the other galaxy and form lensing images. All of the subhalos with enough mass inside the ``main deflector" or along the line-of-sight will perturb these images. By studying these data sets, we can place constraints on the subhalo mass function in the corresponding range. The current observational limit lies around $10^9 M_{\odot}$~\cite{Hsueh_2019} with new proposed methods using machine learning to potentially push the resolution down to $10^7-10^8 M_{\odot}$~\cite{_eng_l_2023, tsang2024substructuredetectionrealisticstrong}. Resolving an independent subhalo, however, would be much more difficult. Currently, researchers have only identified subhalos down to $10^9M_{\odot}$~\cite{Hezaveh_2016}. With future data from VLBI \cite{kadler2024collectiongermanscienceinterests} and JWST \citep{Diego_2023}, however, we might be able to probe substructures down to the galaxy formation cutoff. 

For a subhalo with an even smaller mass down to $\sim 10^6 M_{\odot}$, it has been proposed that one can look for gaps left in the stellar stream caused by their perturbation~\cite{2002MNRAS.332..915I,2002ApJ...570..656J,2012ApJ...748...20C,10.1093/mnras/stw1957}. For example, analysis of the \textit{Gaia} data \citep{Gaia2016, 2018} suggests that the GD-1 stream \citep{Grillmair_2006, Price_Whelan_2018} might be perturbed by a mass between $10^6-10^8 M_{\odot}$~\cite{Bonaca_2019}, with recent studies argue that a self-interacting DM (SIDM)~\cite{Spergel_2000, TULIN20181, adhikari2022astrophysicaltestsdarkmatter} subhalo might actually be preferred over a CDM subhalo as the main perturber ~\cite{zhang2024gd1stellarstreamperturber}. The launch of the Vera C. Rubin Observatory Legacy Survey of Space and Time (LSST)~\cite{ivezić2018lsstsciencedriversreference, drlicawagner2019probingfundamentalnaturedark}, is expected to significantly enhance the precision with which these gaps are measured, and therefore aid in probing even lighter subhalos. However, any astronomical traces left by smaller subhalos than $10^6 M_{\odot}$ would currently be beyond the resolution capabilities of existing technologies. Nonetheless, if these subhalos happen to fly past \textit{a detector on Earth}, the local DM density would increase by orders of magnitude. This increase would significantly enhance the DM probability of detection. Should these attempts result in a non-detection, it would translate to tighter constraints on the DM parameter space than previously thought. Consequently, it is imperative to investigate the local abundance of such light subhalos and assess their potential impact on direct detection experiments. Unfortunately, due to the sparse distribution of these small dark subhalos, the expected number of encounter events per year, as we will show below in Eq.~\ref{eq:rate}, will be at the order of $10^{-8}$, making it highly improbable that such an event will be observed in direct detection experiments.  

A paleo-detector is a proposed alternative to direct detection that can also track these tiny subhalos \cite{PhysRev.133.A1443, doi:10.1126/science.149.3682.383, annurev:/content/journals/10.1146/annurev.ns.15.120165.000245, GUO2012233}. In these detectors, the passage of DM through the Earth can leave tell-tale tracks in ancient natural minerals found deep underground. This approach allows researchers not only to investigate the nature of DM, but also to search for low-energy neutrinos and cosmic rays. These particles can provide clues about various astrophysical events, including the formation history of the solar system and core collapse supernovae~\cite{Drukier_2019, Baum_2020, Baum_2020_803, PhysRevD.104.123015, instruments5020021}. The primary advantage of the paleo-detector is its remarkably long exposure time, spanning approximately $\mathcal{O}(10^9)$ years.  The long exposure time makes up for the low probability of a subhalo encounter event and gives it an advantage over the traditional direct detection experiments, making it potentially an ideal system to look for substructures. Previous studies have shown that Paleo-detectors can serve as a competing DM detector and can even surpass our current direct detection limit in the mass range $m_{\rm{DM}} \sim 1-10$ GeV~\cite{instruments5020021}. It has also been shown that a time series of Paleo-detectors can be used to constrain some of the parameters for substructures like dark disks, or subhalos in the mass range $10^4-10^8 M_{\odot}$~\cite{PhysRevD.104.123015}. 

In this study, we further build on this method that uses a time series of Paleo-detectors to study subhalos, and explore Paleo-detectors' expected detection ability over a wider mass range on the mass concentration relation for different DM parameters. We find that, even though the yearly encounter rate is small, Paleo-detectors are able to set meaningful constraints on the mass-concentration relation for some of the DM models not yet excluded. Additionally, we will briefly discuss the potential interesting physical applications of Paleo-detectors in some non-CDM models like strongly interacting DM subcomponent and atomic DM (aDM).

This paper is structured as follows: In Sec.\ref{sec:Review}, we will first briefly review Paleo-detectors. Then, in Sec.\ref{sec:ER}, we will describe how we estimate the expected number of encounter events of dark subhalos in such detectors. Specifically, in Sec.\ref{sec:LDN} and Sec.\ref{sec:ECS}, we explain how to compute the local differential number density from previous studies, and calculate the encounter cross section of a subhalo passing through the Earth. In Sec.\ref{sec:DER}, we put everything together to get a differential encounter rate and give an estimate of encounter events per year over a wide mass range of subhalos. In Sec.~\ref{sec:IDD}, we examine the anticipated signal detection by a Paleo-detector from expected encounters, and establish constraints on the mass-concentration relation. Finally, we further discuss future perspectives for some non-CDM models in Sec.\ref{sec:outlook} and conclude in Sec.\ref{sec:conclusion}.

\section{\label{sec:Review}Brief Review of Paleo-Detectors}

As DM particles (or other ionizing radiation) pass through a mineral, they can interact with the nuclei or electrons in the crystal lattice. This interaction results in the DM particle transferring energy to the atoms in the mineral. The energy transferred during the interaction displaces atoms from their lattice positions, creating a trail or ``track" of damage within the crystal structure that can be preserved in the mineral for geological timescales~\cite{PhysRev.133.A1443, doi:10.1126/science.149.3682.383, annurev:/content/journals/10.1146/annurev.ns.15.120165.000245, GUO2012233}. Therefore, the observable of a Paleo-detector is the number of tracks of a given length, the track length spectra, whose distribution is sensitive to different types of incoming particles. In this section, we will briefly go through how these track length spectra are computed. More detailed discussions of the spectrum and readout processes can be found in Refs.~\cite{Drukier_2019, Baum_2020, Baum_2020_803, PhysRevD.104.123015, instruments5020021}, along with discussions of background from neutrinos and radiative elements, and mineral selections. 

A recoiling nucleus in Paleo-detectors will leave a damage track. The length of a damage track, $x_t$, for a recoiling nucleus with deposited energy $E_R$ can be estimated as
\begin{equation}\label{eq1}
    x_{t}(E_{R})=\int^{E_R}_0 \left|\frac{dE}{dx_t}\right|^{-1}dE, 
\end{equation}
where $|\frac{dE}{dx_T}|$ is the stopping power of the nucleus in the target material, characterizing how much energy of the recoiling nucleus is lost per unit length traveled in the material. The real length of the trajectory may differ from this value, as the trajectory might either not be straight, or only part of it is recorded as a damage track. However, previous studies \cite[see e.g.][]{Drukier_2019} have stated that these effects are small. Concerns may also arise regarding the potential impact of thermal annealing over geological timescales, which can indeed be substantial. However, it is important to note that its effect on both the signal (i.e. DM interaction) and background (i.e. cosmic rays or other non-DM processes) is likely to be comparable. So the signal to background ratio would roughly stay the same. 

The track spectrum is constructed as the number of tracks for each track length $x_t$. So, in order to obtain a spectrum, we need an expression for the differential recoil event rate per length, which we integrate over the exposure time. It is standard to use the differential event rate per deposited energy per unit mass of the material $(dR/dE_R)_k$ for a given type of nucleus $k$ (similarly to direct detection experiments). This quantity indicates the expected number of events originating from scattering events with a specific type of nucleus per unit time, given a recoil energy $E_R$, within a unit mass of the target material. We can couple this differential event rate with the stopping power, and sum over all species of nuclei (i.e. sum over $k$), and the mass fraction $\xi_k$, to reach the desired expression,
\begin{equation}
    \frac{dR}{dx_t}=\sum_k \xi_k \left(\frac{dR}{dE_R}\right)_k\left(\frac{dE_R}{dx_t}\right)_k.
\end{equation}
In practice, however, the precision of observations is constrained by the resolution of the readout process \cite[see e.g.][]{Drukier_2019}. Consequently, it is necessary to convolve the true spectrum with a window function to account for these limitations. Then, for the $i$-th bin with an observed track length $x_t \in [x_i^{\rm{\rmmin}}, x_i^{\rm{max}}]$, the event rate per unit target mass is given by
\begin{equation}
    R_i(x_i^{\rmmin}, x_i^{\max})=\int^{\infty}_0 W(x'_t; x_i^{\rmmin}, x_i^{\max})\frac{dR}{dx'_t}dx'_t,
\end{equation}
where $W$ is a window function which we assume to be Gaussian-distributed, with the true length $x_t'$ as its mean and the resolution $\sigma_{x_t}$ as its variance. It can therefore be written as
\begin{equation}
    W=\frac{1}{2}\left[\text{erf}\left(\frac{x_t'-x_i^{\rmmin}}{\sqrt{2}\sigma_{x_t}}\right)-\text{erf}\left(\frac{x_t'-x_i^{\max}}{\sqrt{2}\sigma_{x_t}}\right)\right].
\end{equation}
With all of these aforementioned steps, we can now write the number of tracks in each length bin as
\begin{equation}
    N_i=M_s\int^T_0R_idt, 
\end{equation}
where $M_s$ is the sample mass and $T$ is the integration time.

In this study, we will work with a ``time series of Paleo-detectors." This is a series of Paleo-detectors that are spaced in time. The samples were formed at different ages, and thus have varying integration times. Each sample will have a track spectrum from Milky Way (MW) DM interactions, with the number of tracks linear in time. For a given time-dependent event, for example a subhalo encounter, any sample older than the time of the event will record an extra contribution from such an event in addition to the MW spectrum. By comparing their track spectra, we can study the time-dependence of the DM interactions.

\section{\label{sec:ER}Subhalo Encounter Rate}

In order to estimate how the low mass dark subhalos affect DM direct detection on Earth, we first characterize the yearly encounter rate of such dark subhalos with detection experiments. 

We set up the cross section $\sigma(v_0, M_{\rm_{\sub}})$ within which a subhalo of mass $M_{\rm{\sub}}$, with initial velocity $v_0$ can hit Earth and affect direct detection experiments. Then, the incoming rate of subhalos with a velocity $v_0$, velocity distribution $f(v_0)$, and number density $n$ can be given by $f(v_0)\sigma(v_0, M_{\sub})v_0n dv_0$. The differential encounter rate per unit subhalo mass is therefore
\begin{equation}\label{eq6}
    \frac{dR_{\sub}}{dM_{\sub}}=\int \frac{dn_{\rm{\sub}}}{dM_{\sub}}~f(v_0)~v_0~\sigma(v_0, M_{\rm{\sub}})~dv_0, 
\end{equation}
where $dn_{\rm{\sub}}/dM_{\sub}$ is the local differential number density of the dark subhalos.

We assume the velocity distribution of the dark subhalos follows that of the Standard Halo Model~\cite{LEWIN199687, RevModPhys.85.1561}, and is of the form
\begin{equation}
    \begin{split}
            f_{\rm{SHM}}(v|\sigma_0, v_{\obs})=&\frac{v}{\sqrt{2 \pi}\sigma_0 v_{\obs}}e^{-(v+v_{\obs})^2/2 \sigma_0^2}\\
            &\times (e^{2vv_{\obs}/\sigma_0^2}-1),
    \end{split} 
\end{equation}
where $\sigma_0 \approx 166$~km/s is the velocity dispersion and $v_{\obs} \approx 232$~km/s is the speed of the Sun relative to the halo rest frame~\cite{10.1093/mnras/stw1957, RevModPhys.85.1561, 10.1093/mnras/stae034}. More detailed modeling of the velocity distribution can be performed as in~\cite{Necib_2019, Necib_2019_01}. We have checked that the detailed choice of the velocity distribution does not affect the estimation of the rate. Thus, we are now only left with the local differential number density $dn_{\rm{\sub}}/dM_{\rm{\sub}}$ and the cross section $\sigma(v_0, M_{\sub})$ to be discussed in the following subsections. 

\subsection{\label{sec:LDN}Local Differential Number Density}

The expression that we employ for the local differential number density $dn_{\rm{\sub}}/dM_{\rm{\sub}}$ of the DM subhalo is given in~\cite{10.1093/mnras/stw1957}, which we discuss below. It found broadly similar local differential number density as those in the public catalogs of the simulation Via Lactea II~\cite{2008Natur.454..735D} for subhalos within 50~\rm{kpc} of the center of the halo. 

In order to estimate the differential number density in the solar neighborhood, Ref.~\cite{10.1093/mnras/stw1957} worked through its scaling with the distance to the center of the halo $r$, and the mass of the subhalos $M_{\sub}$ by taking the simulation results from the Aquarius Project~\cite{2008MNRAS.391.1685S} as follows. Assuming the distance and mass scalings are independent of each other,\footnote{In other words, we are ignoring effects like subhalos are more tidally disrupted toward the center of the galaxy and more massive subhalos can survive longer from the tidal disruption; we are assuming the mass function is constant as a function of $r$ from the center. } one can infer the distance scaling from the number density profile $n_{\sub}$ and the mass scaling from the mass function\footnote{$N_{\sub}$ and $n_{\sub}$ are different. $N_{\sub}$ is the number of subhalos as it pertains to the mass function $dN_{\sub}/dM_{\rm{\sub}}$, while $n_{\sub}$ is the number density of subhalos.} $dN_{\sub}/dM_{\rm{\sub}}$. The number density profile of subhalos in a Milky Way-like analog is found to be well described by an Einasto profile \cite{1965TrAlm...5...87E} as studied in the Aquarius Project~\cite{2008MNRAS.391.1685S} 
\begin{equation} \label{eq:nsub}
    n_{\rm{\sub}}\propto \exp \left[-\frac{2}{\alpha}\left(\left(\frac{r}{r_{-2}}\right)^{\alpha}-1\right) \right]
\end{equation}
with $\alpha=0.678$ and $r_{-2}=0.81 \times~r_{200}=199$~kpc, which is defined as the radius where the slope of the density profile is $-2$. The virial radius $r_{200}$ is defined as the radius within which the average density of the system equals the critical density of the universe $\rho_c=3H(t)^2/(8\pi G)$ multiplied by an overdensity factor $\Delta$ which is often chosen as 200. The virial mass $M_{200}$ is defined as the mass contained within the virial radius. $H(t)$ is the Hubble constant at time $t$ when we calculate the virial radius and $G$ is the gravitational constant. This expression is found to be valid across a mass range from $10^5M_{\odot}$ to $10^{10}M_{\odot}$ ~\cite{2008MNRAS.391.1685S}. 

The inferred mass of the MW in the literature spans a wide range of masses, from as low as $0.5 \times 10^{12}M_{\odot}$ up to as high as $3\times 10^{12}M_{\odot}$~\cite{10.1111/j.1365-2966.2007.12748.x, 10.1111/j.1365-2966.2010.16708.x, 10.1111/j.1365-2966.2010.16774.x, van_der_Marel_2012, Watkins_2019, 10.1093/mnras/stae034}.  For our purposes, however, we will assume the MW has a mass of $\sim 10^{12}M_{\odot}$ since different choices of mass will not significantly change the result. Given that the galaxy simulated in the Aquarius Project has a virial mass of $M_{200}=1.839\times 10^{12} M_{\odot}$, for self-consistency, we scale down the mass and distances appropriately. The distance scales as $\sim M^{1/3}$ assuming the same halo concentration and $\alpha$. So $r_{-2}$ is scaled to 162.4~kpc with other fitting parameters unchanged.

To get $dn_{\rm{\sub}}/dM_{\rm{\sub}}$, we need to combine Eq.~\ref{eq:nsub} with the subhalo mass function given in Ref.~\cite{2008MNRAS.391.1685S}. The latter is defined as
\begin{equation} \label{eq:subhalo_mass_function}
    \frac{dN_{\sub}}{dM_{\rm{\sub}}}=a_0 \left(\frac{M_{\rm{\sub}}}{m_0}\right)^{\beta},
\end{equation}
where $a_0=3.62\times 10^{-5}M_{\odot}^{-1}$, $m_0=2.52\times 10^7M_{\odot}$, and $\beta=-1.9$. Eq.~\ref{eq:subhalo_mass_function} is valid within $r_{50}=433$~kpc as studied in Ref.~\cite{2008MNRAS.391.1685S}, which encloses a mean density of 50 times of the critical density,\footnote{The critical density is found to be also roughly 200 times of the \textit{background density}, which is why it is used as an alternative definition to the virial radius in some studies. } and therefore valid for studying the local density near the solar neighborhood. Scaling down both $a_0$ and $r_{50}$ similarly as before, we get $a_0=1.77\times 10^{-5}M_{\odot}^{-1}$ within $r_{50}=353$~kpc. Now that we know how the number density varies as a function of the radial distance from the Galactic Center, and how the mass function scales with subhalo mass, we can factor them out with an extra overall normalization factor to get a well normalized local differential number density
\begin{equation}\label{dn}
    \frac{dn_{\rm{\sub}}}{dM_{\rm{\sub}}}=c_0 \left(\frac{M_{\rm{\sub}}}{m_0} \right)^\beta \exp\left[-\frac{2}{\alpha} \left(\left(\frac{r}{r_{-2}}\right)^{\alpha}-1 \right)\right],
\end{equation}
with the normalization factor $c_0=2.02\times 10^{-13}M_{\odot}^{-1}$~kpc$^{-3}$ as found in Ref.~\cite{10.1093/mnras/stw1957}. We note, however, that the aforementioned expression and cited values are all based on a collisionless N-body simulation in the Aquarius Project that neglected the effects of baryons~\cite{10.1093/mnras/stw1957,2008MNRAS.391.1685S}. Ref.~\cite{D'Onghia_2010} found using \textsc{GADGET3} \cite{2005MNRAS.364.1105S}, that the presence of a baryonic disk that makes up $10\%$ of the total mass of the host galaxy decreases the abundance of the subhalos at masses of $10^7 M_{\odot}$ by a factor of 3~\cite{D'Onghia_2010}. We therefore account for the presence of baryons similarly to Ref.~\cite{10.1093/mnras/stw1957} by scaling down the number of subhalos by an effective extra factor of 3. Assuming the validity of this expression across all masses of interest, we are now ready to apply it to compute the local number density in any mass range.

\subsection{\label{sec:ECS}Encounter Cross Section}

\begin{figure}[t]
\begin{center}
  \includegraphics[width=0.8\linewidth]{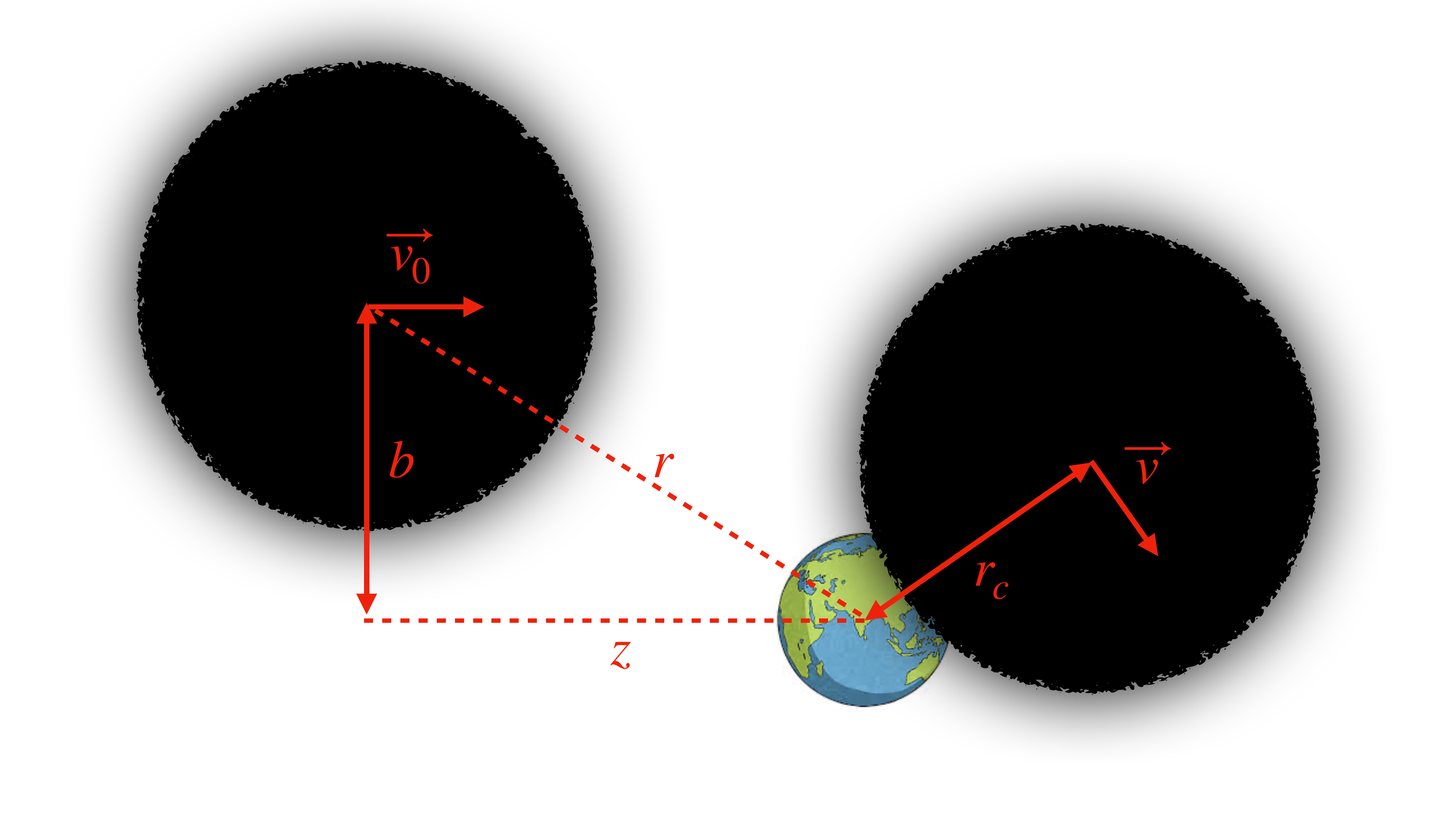}
 \caption{Illustration of a subhalo encountering event. Gravitational effect is exaggerated. }
  \label{fig:ppt}
 \end{center}
 \end{figure}
 
Next, we study the encounter cross section between a dark subhalo and Earth, where their interaction is purely gravitational. Using conservation of energy and angular momentum, and letting $b$ be the impact parameter and $r_c$ be the closest distance from the Earth to the center of the subhalo as illustrated in Fig.~\ref{fig:ppt}, we can solve for the largest possible impact parameter $b_{\rm{max}}$ that leads to an impact on the local DM density, and therefore its detection,
\begin{equation}\label{bmax}
b_{\rm{max}}=\sqrt{1+\frac{2GM_E}{v_0^2 r_c}}\, r_c
\end{equation}
 with $M_E$ being the mass of the Earth, and $v_0$ the initial velocity of the subhalo. Then the scattering cross section is given by integrating over all angles and all impact parameters up to $b_{\rm{max}}$
\begin{equation} \label{eq:sigma}
\sigma(v_0, M_{\rm_{\sub}})=\int_{0}^{2\pi}d\phi\int_0^{b_{\rm{max}}}bdb=\pi b_{\rm{max}}^2.
\end{equation}
where the mass dependence on $M_{\sub}$ is found in $r_c$ (and therefore $b_{\rmmax}$) implicitly.
\begin{figure}[t]
\begin{center}
  \includegraphics[width=0.95\linewidth]{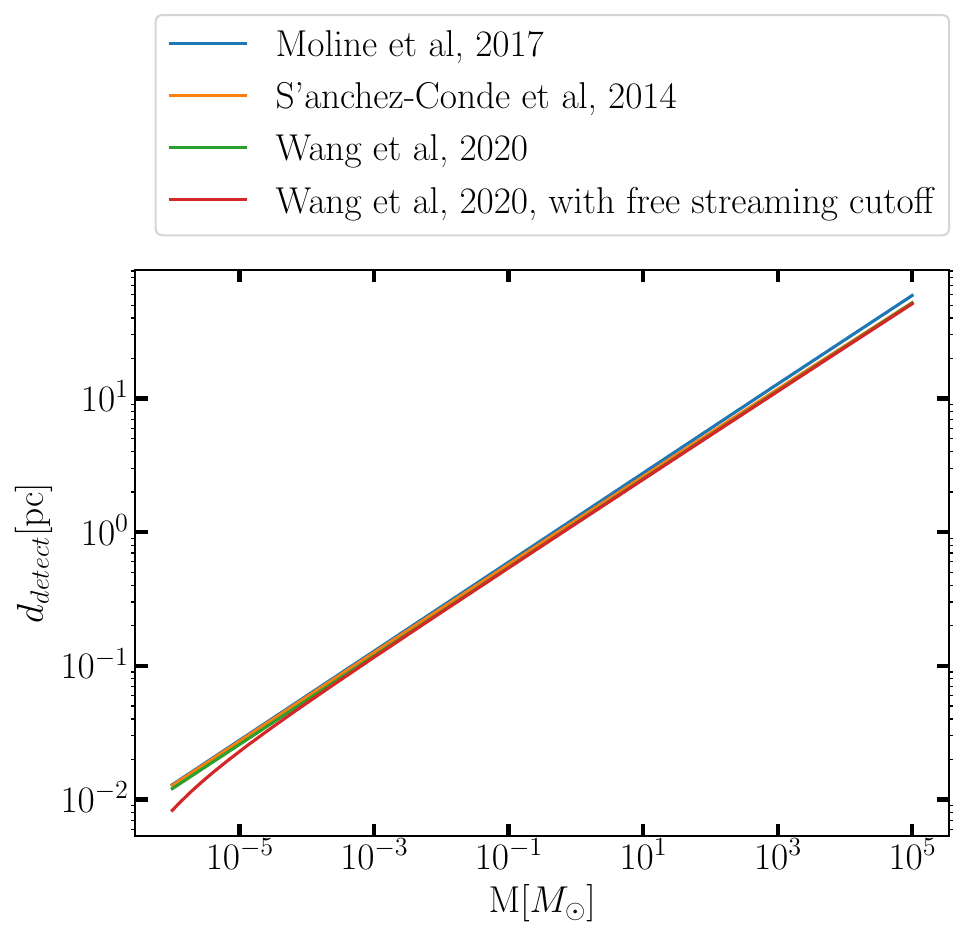}
 \caption{$\ddedect$ over the mass range of our consideration with four different concentration models. $c(M)$ proposed by Moline et al, S'anchez-Conde et al, and Wang et al are in agreement over the entire mass range, except for Wang et al but with free-streaming cutoff at the lower end. The radius increases monotonically with the mass as we would naively expect.}
  \label{fig:ddetect}
 \end{center}
 \end{figure}

The goal is to find $r_c$, the closest distance from Earth to the center of the subhalo, that is most relevant to detection experiments. We need such a radius at which the dark subhalo's density is higher than the average DM density in the solar neighborhood, which is around $0.3~$GeV/cm$^3$~\cite{Read_2014}. The density profile of a DM halo is usually modeled as the Navarro–Frenk–White (NFW) profile~\cite{Navarro_1996, Navarro_1997}, which is given by
\begin{equation}
        \rho(x=r/r_{200})=\frac{\rho_{200}}{3A_{\NFW}x(c^{-1}+x)^2},
\end{equation}
where $x=r/r_{200}$ is the fractional distance, $\rho_{200}$ is the average density enclosed within the virial radius $r_{200}$, and $A_{\NFW}$ is a numerical factor determined by the halo's concentration $c$ as 
\begin{equation*}
A_{\NFW}=\ln(1+c)-\frac{c}{1+c}.    
\end{equation*}
While a higher density region would be more detectable, it would correspond to a smaller inner region and thus an even shorter exposure time. So in order to get the most optimistic estimate, we should integrate out to as far as possible and thus to a radius that encloses all the region with density higher than the local DM density. The critical density is given by $\rho_{200}=200\rho_c=200\times \frac{3H(t)^2}{8\pi G}\sim 0.001~\rm{GeV}/\rm{cm}^3$, which is two orders of magnitude smaller than the local density. We therefore require the density of DM in the subhalo to be at least two orders of magnitudes higher than $\rho_{200}$, so that it is comparable to the local density of DM in the solar neighborhood, and will cause an $\mathcal{O}(1)$ change in the expected signal. Thus we define $\ddedect$, which is the radius where the density is around 20000 times of the critical density and around the local density in the solar neighborhood. 

Since the average density within the virial radius $r_{200}$ is a constant, $r_{200}$ is then only a function of subhalo mass $M_{\sub}$, and thus the fractional distance $x$ is a function of distance $r$ and $M_{\sub}$. We should also notice $A_{\NFW}$ is only a function of concentration $c$. Then if we relate the concentration $c$ to the subhalo mass $M_{\sub}$ using the mass-concentration relation, then $A_{\NFW}$ is also only a function of mass, and so is $\ddedect$. Hence, we find the dependence of $\ddedect$ versus $M_{\sub}$, which we show in Fig.~\ref{fig:ddetect}. 

In Fig.~\ref{fig:ddetect}, we inspect four different analytical mass-concentration relation models~\cite{S_nchez_Conde_2014, Molin__2017, Wang_2020} as shown in four different colors: 
First, in ~\cite{S_nchez_Conde_2014}, a set of median concentrations and halo masses available in the literature from different simulations is summarized across the mass range from $10^{15}M_{\odot}$ down to as low as the Earth mass $10^{-6}M_{\odot}$, and a mass-concentration model $c(M)$ model is fitted using these data points. 
Second, in ~\cite{Molin__2017}, another $c(M)$ is established using Via Lactea II \citep{2008Natur.454..735D} and ELVIS \citep{2014MNRAS.438.2578G} for subhalos with masses $10^6-10^{11} M_{\odot}$. 
Third, and most recently, in ~\cite{Wang_2020}, a multi-zoom technique is used to conduct a cosmological simulation over the full mass range from $10^{15}M_{\odot}$ to $10^{-6}M_{\odot}$. A $c(M)$ is given for results with and without a free-streaming cutoff at small spatial scales. Free-streaming happens in the early universe when DM particles move freely, and therefore can smooth out small scale structures \citep{ZYBIN1999262, PhysRevD.64.083507}. The cutoff scale of such streaming is model-dependent, and is typically much smaller than the Earth's mass (in some cases, it can even get close to the DM particle mass \citep{Bringmann_2009, Marsh_2016}). For illustrative purpose, however, a thermal WIMP mass of 100~GeV is chosen, so that the turnaround point of the cutoff happens around the Earth mass at the lower end of the subhalo mass spectrum. We should note that the subhalos in ~\cite{Wang_2020} are too low in density to represent MW subhalos since they are from voids. The tidal radius of these subhalos near the location of the Sun is smaller than the scale radius of the subhalo, which suggests it would be disrupted. Thus, the actual concentration should be higher and the $c(M)$ relation from like ~\cite{Molin__2017} might be more accurate. 

We find that $\ddedect$ does not vary much for different $c(M)$ models except for the one with free-streaming cutoff at the lower end (see Fig.~\ref{fig:ddetect}). The radius increases monotonically with the mass as we would naively expect. Additionally, in Eq.~\ref{bmax}, the term $2GM_E/v_0^2 \ddedect \ll 1$ for $v_0 \sim \mathcal{O}$(100~km/s), and $\ddedect \sim \mathcal{O}$(1~pc).

We therefore we can approximate the impact parameter $b_{\rmmax}$ as $\ddedect$.  

\subsection{\label{sec:DER}Differential Encounter Rate}

\begin{figure}[t]
\begin{center}
  \includegraphics[width=0.95\linewidth]{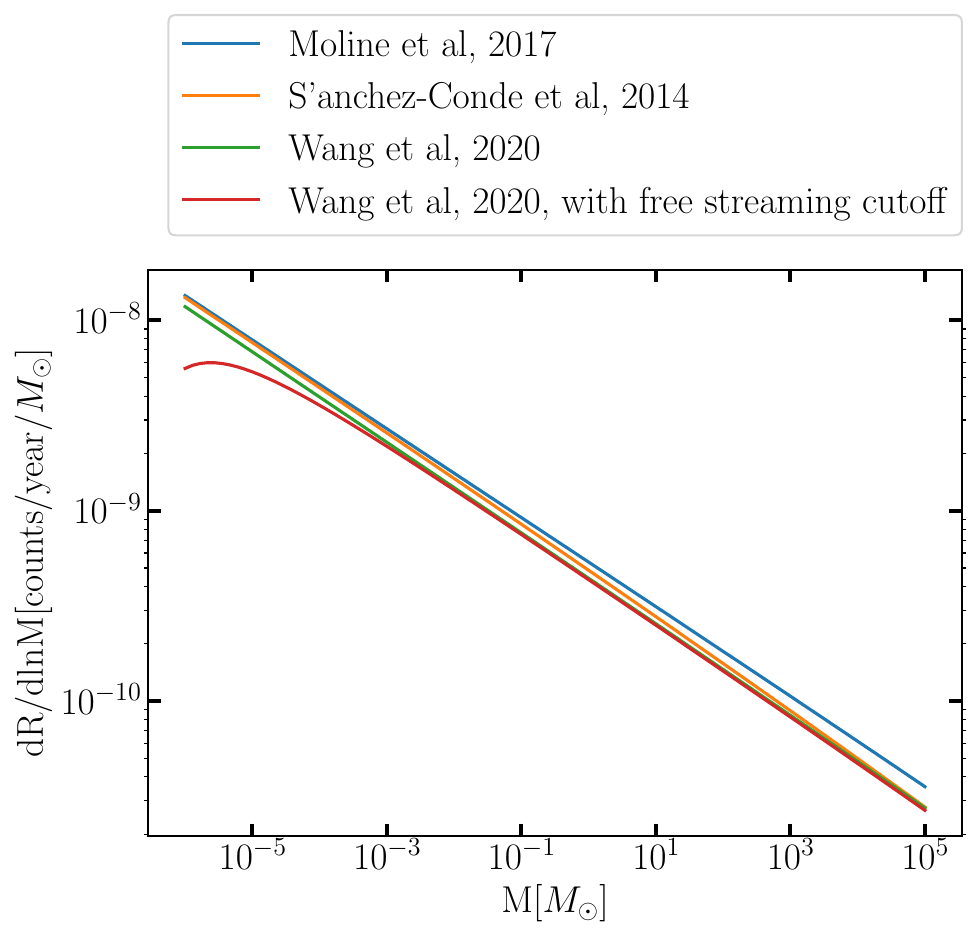}
 \caption{The differential event rate over the mass range of our consideration with four different concentration models. There is negligible variation for different $c(M)$ models except for the one with a free-streaming cutoff at the lower end.}
  \label{fig:dR}
 \end{center}
 \end{figure}

Putting all the pieces of the differential encounter rate together defined in Eq.~\ref{eq6}, having included the differential number density in Eq.~\ref{dn}, and the scattering cross section in Eq.~\ref{eq:sigma}, we show in Fig.~\ref{fig:dR} the differential encounter rate for the different $c(M)$ models discussed above. To get the total encouter rate, we integrate $\frac{dR_{\sub}}{dM_{\sub}}$.
The lower limit of the integral is set as $10^{-6}M_{\odot}$, which is determined by the free-streaming cutoff in Ref.~\citep{Wang_2020}\footnote{The scaling of the differential rate in terms of the subhalo mass is $\frac{dR_{\sub}}{dM_{\sub}}\sim M_{\sub}^{-1.6}$ (A more detailed discussion on the scaling is presented in Appendix.~\ref{app:scaling})}. 
The upper limit of the integral is chosen as $10^{5}M_{\odot}$. This choice minimally affects the final answer.

The total encounter rate of a subhalo within the mass range studied is therefore
\begin{equation} \label{eq:rate}
\int_{M_{\rm{\rmmin}}=10^{-6}M_{\odot}}^{M_{\rm{max}}=10^5 M_{\odot}}\frac{dR_{\sub}}{dM_{\sub}}dM_{\sub} \sim 10^{-8}/{\rm{year}}.
\end{equation}

\section{\label{sec:IDD}Results}

Even though the likelihood of a yearly encounter event with a subhalo is $\sim 10^{-8}$, an extremely rare event to occur for a direct detection experiment, one can compensate for the small rate with a long enough integration time using a Paleo-detector \cite{PhysRev.133.A1443, doi:10.1126/science.149.3682.383, annurev:/content/journals/10.1146/annurev.ns.15.120165.000245, GUO2012233}. Thus, the expected number of encounters, over the age of the Earth ($\mathcal{O}(10^9)$ years), would be of order unity. 

In the case of such an encounter, we can then use Paleo-detectors to look for the subhalo and constrain its properties. We will treat the spectrum of tracks left by DM particles from the MW halo as background, and consider the spectrum of tracks left by subhalo DM particles as the signal. We will use a likelihood ratio test to distinguish a time-varying subhalo signal from a time-independent MW background. 
In this section, we will first calculate the excess signal coming from subhalo events compared to those of the MW halo, followed by a discussion of the possible constraints set on the mass-concentration relation for a time series of Paleo-detectors. The code we used for calculating the track spectrum and sensitivity analysis is modified from the code published in \cite{PhysRev.133.A1443, doi:10.1126/science.149.3682.383, annurev:/content/journals/10.1146/annurev.ns.15.120165.000245, GUO2012233}. 

\subsection{\label{sec:DD}Signal/Background Rates}

\subsubsection{\label{sec:DD}Milky Way halo}

For spin-independent interactions, the differential nuclear recoil rate per unit target mass for a WIMP particle with mass $m_{\chi}$, scattering off a nuclei with atomic number $A$ and deposited energy $E_R$ is given by \cite{PhysRev.133.A1443, doi:10.1126/science.149.3682.383, annurev:/content/journals/10.1146/annurev.ns.15.120165.000245, GUO2012233}
\begin{equation}\label{dRMW}
 \frac{dR}{dE_R}=\frac{A^2F(E_R)^2\sigma^{SI}}{2 \mu^2_{p \chi}m_{\chi}}\rho_{\chi} \eta_{\chi}(v_{\rm{\rmmin}})
\end{equation}
where $F(E_R)$ is the nuclear form factor that captures the internal nuclear structure of the recoiling nucleus, assumed to be the Helm parametrization \cite{PhysRev.104.1466,LEWIN199687,D_da_2007}, $\sigma^{SI}$ is the spin-independent DM-nucleon interaction cross-section, and $\mu_p=m_{\chi} m_p/(m_{\chi}+m_p)$ is the reduced mass. The last term $\eta_{\chi}(v_{\rmmin})$ is the mean inverse speed, which is given by,
\begin{equation}\label{inverse}
\eta_{\chi}(v_{\rm{\rmmin}})=\int_{v>v_{\rm{\rmmin}}}\frac{f(\bold{v})}{v}d^3\bold{v}
\end{equation}
where we assume the general 3-D Standard Halo Model/Maxwellian velocity distribution. The distribution $f(\bold{v})$ is assumed to have a velocity dispersion $\sigma_v=166$~km/s, and boosted to the helio-frame by $v_{\odot}=248$~km/s \citep{PhysRevD.104.123015}. The lower bound $v_{\rmmin}=\sqrt{m_N E_R/(2(\mu_{\chi})^2)}$ is limited by the energy threshold $E_R$ of the detector \citep{PhysRevD.104.123015}. 

\subsubsection{\label{sec:DD}Subhalo}

Unlike a MW halo event, which is accumulated continuously over time, a subhalo signal is a one-time event over a relatively short integration period. We are therefore interested in the total differential number of recoils within that period between times $t_0$ and $t_1$, $dn_R/dE_R=\int^{t_1}_{t0} dR/dE_R dt$. For any subhalo encounter event with a given mass $M_{\sub}$, the differential number of recoils can typically be written as \citep{PhysRevD.104.123015}, 
\begin{equation}\label{shsignal}
\begin{split}
\left( \frac{dn_R}{dE_R}\right)^{\sub}&=\frac{A^2F^2\sigma^{SI}}{2 \mu^2_p m_{\chi}} \\
&\times \frac{1}{v_{\sub}} \int^{z_1}_{z_0}\rho^{\sub}_{\chi}(r, M_{\sub}) \eta^{\sub}_{\chi}(r;v_{\rmmin})dz
\end{split}
\end{equation}
where we have neglected the gravitational effect from the Earth since the mass range we are considering is in general comparable or greater than the Earth's mass. $z$ denotes the distance the Earth has traveled through in the subhalo and $r(z)=\sqrt{b^2+z^2}$ is the position of the Earth to the center of the subhalo, and depends on the impact parameter $b$, (see Fig.~\ref{fig:ppt}). 

Then in order to get an expected signal, we can combine Eq.~\ref{shsignal} with the expected number of subhalo encounter events as in Eq.~\ref{eq6} integrated over our mass range and the age of the sample, which yields
\begin{equation}\label{eq18}
\begin{split}
\left( \frac{dn_R}{dE_R}\right)^{\sub}_{\rm{exp}}&=\frac{A^2F^2\sigma^{SI}}{2 \mu^2_p m_{\chi}}\times T\int^{M_{\rmmax}}_{M_{\rmmin}} dM_{\sub} \frac{dn_{\sub}(M_{\sub})}{dM_{\sub}} \\
&\times \int^{b_{\rmmax}}_0 2\pi b db \int f(\bold{v_{\sub}})d^3\bold{v_{\sub}} \\
&\times \int^{z_1(b)}_{z_0(b)}\rho^{\sub}_{\chi}(r, M_{\sub}) \eta^{\sub}_{\chi}(r;v_{\rmmin})dz
\end{split}
\end{equation}
where $T$ is the age of the sample. The mass is integrated in the range between $M_{\rm{\rmmin}} = 10^{-6}M_{\odot}$ and $M_{\rm{max}}=10^5M_{\odot}$. The impact parameter $b$ is integrated up to $b_{\rmmax}$ as in Eq.~\ref{bmax}. 

For simplicity, we assume the encounter to be symmetric with $z_0(b)=-z_1(b)=z_{\rmmax}(b)$. There are several physical scales that we can consider for the furthest distance $z_{\rmmax}$ from the Earth to the center of the subhalo during an encounter, i.e. virial radius $r_{\rm{vir}}$, scale radius $r_s$, or tidal radius $r_t$. We have found, however, that different choices of radii will not drastically change the number of events, as the outer part of the subhalo is less dense and thus contributes far less than the central part to the number of tracks. And, lastly, the mean inverse speed $\eta$ is defined the same as in Eq.~\ref{inverse}, except that its denominator, the relative velocity to the detector, is now the vector sum of the subhalo velocity and the velocity of the DM particle relative to the subhalo $\bold{v_{\rm{rel}}}=|\bold{v_{\sub}}+\bold{v_{DM}}|$.

Writing out the velocity integral part of Eq.~\ref{eq18}, and switching the order of the velocity integrations, we find
\begin{equation}
\begin{split}
&\int^{\infty}_0 f(\bold{v_{\sub}})d^3\bold{v_{\sub}}\eta_{\chi}^{\sub}(v_{\rmmin})=\\
&\int_0^{\infty} f(\bold{v_{\sub}})d^3\bold{v_{\sub}} \int_{|\bold{v_{DM}}|>|\bold{v_{\rmmin}-\bold{v_{\sub}}}|}\frac{f(\bold{v_{DM}})}{|\bold{v_{\sub}}+\bold{v_{DM}}|}d^3\bold{v_{DM}}=\\
&\int_0^{\infty} f(\bold{v_{DM}})d^3\bold{v_{DM}} \int_{|\bold{v_{\sub}}|>|\bold{v_{\rmmin}-\bold{v_{DM}}}|}\frac{f(\bold{v_{\sub}})}{|\bold{v_{\sub}}+\bold{v_{DM}}|}d^3\bold{v_{\sub}}.
\end{split}
\end{equation}
The velocity distribution of subhalos is again assumed to be a Maxwellian velocity distribution with a velocity dispersion $\sigma^{\sub}_v=166$~km/s. The DM particles velocity distribution within a subhalo is the same as in the last section except that its velocity dispersion here is governed by the subhalo potential and thus depends on the distance to the center of the subhalo. This is an intertwined quintuple integral which is computationally expensive. 

We therefore make a few approximations to reduce the number of integrations. First, given that the subhalo velocity centers around a few hundred km/s and the DM velocity dispersion within a subhalo, albeit position-dependent, is typically smaller than 1~km/s at the masses considered in this work, we can neglect the effect of the latter velocity dispersion. Additionally, the minimal detectable velocity $v_{\rmmin}$ of the detector, which depends on the mass of the nuclei, the mass of the DM particle, and the resolution of the detector, is at the order of 10~km/s and thus much larger than the DM velocity dispersion. We can then approximate $|\bold{v_{\rmmin}-\bold{v_{DM}}}| \approx |\bold{v_{\rmmin}}|$ and $|\bold{v_{\sub}}+\bold{v_{DM}}| \approx |\bold{v_{\sub}}|$. Therefore, the DM velocity distribution can be integrated out. The remaining velocity distribution integral thus becomes independent of other variables and can be computed separately as
\begin{equation}\label{eq20}
\int^{\infty}_0 f(\bold{v_{\sub}})d^3\bold{v_{\sub}}\eta_{\chi}^{\sub}(v_{\rmmin})=\int_{|\bold{v_{\sub}}|>|\bold{v_{\rmmin}}|}\frac{f(\bold{v_{\sub}})}{\bold{v_{\sub}}}\bold{dv_{\sub}}.
\end{equation}

This simplifies the quintuple integral in Eq.~\ref{eq18} to a triple integral, which is much more manageable.

\subsection{\label{sec:PD}Mass-Concentration Relations}

Even though the subhalo contribution to the track spectrum is small, it remains possible to derive constraints on the mass-concentration relation using a time series of Paleo-detectors. 

As discussed in Sec.~\ref{sec:ECS}, the NFW density profile of a subhalo is contingent upon its mass and concentration parameters. Establishing a relationship between concentration and mass allows the density profile to be characterized exclusively by the mass. Despite its importance, there are still no solid arguments on the dependence of concentration on mass, although it is generally believed that the concentration of a subhalo depends on its formation history, such as accretion or tidal disruption, which correlates with its mass \citep{2012MNRAS.427.1322L, Correa_2015, Klypin_2016}. So, constraining the mass-concentration relation could provide valuable insights into the subhalo density profile modeling and formation history. 

In order to constrain the mass-concentration relation, we need to scan over the mass-concentration parameter space while fixing other parameters that can affect the relative size of the subhalo signal compared to the MW halo signal. The main method to calculate the sensitivity region in the mass-concentration plane is to use the subhalo mass $M_{\sub}$ as the parameter that governs the shape of the subhalo signal, and the concentration parameter $c$ as the independent variable. For each value of the concentration, we compute the smallest $M_{\sub}$ where the combined model (sum of the time-varying signal from subhalo, MW signal as a background, and other non-DM background events) is preferred over the MW signal and non-DM background model. 

While varying the mass and concentration, we will treat as fixed the following parameters: the impact parameter $b$, the subhalo velocity $v_{\sub}$, and the subhalo encounter time $T^{\rm{sh}}$. The impact parameter $b$ and subhalo velocity $v_{\sub}$ are fixed by their respective expected values: For the impact parameter, we assume the subhalos are distributed uniformly in the solar neighborhood, and thus the chance of finding a subhalo with an impact parameter $b+db$ can be written as $\frac{2\pi b db}{\pi b_{\max}^2}$, where, again, we choose $b_{\rm{max}}$ to be the scale radius. For the subhalo velocity, we calculate the expectation value of the assumed Maxwellian distribution. We assume a time series of $5$ samples with the oldest having an age of 1~Gyr and the neighboring samples are 200~Myrs apart. As shown in \citep{PhysRevD.104.123015}, the sensitivity contour is weakly dependent on the encounter time between 300~Myr and 900~Myr, so we will just assume the encounter happened 500~Myrs ago. 

\begin{figure}[t]
\begin{center}
  \includegraphics[width=0.95\linewidth]{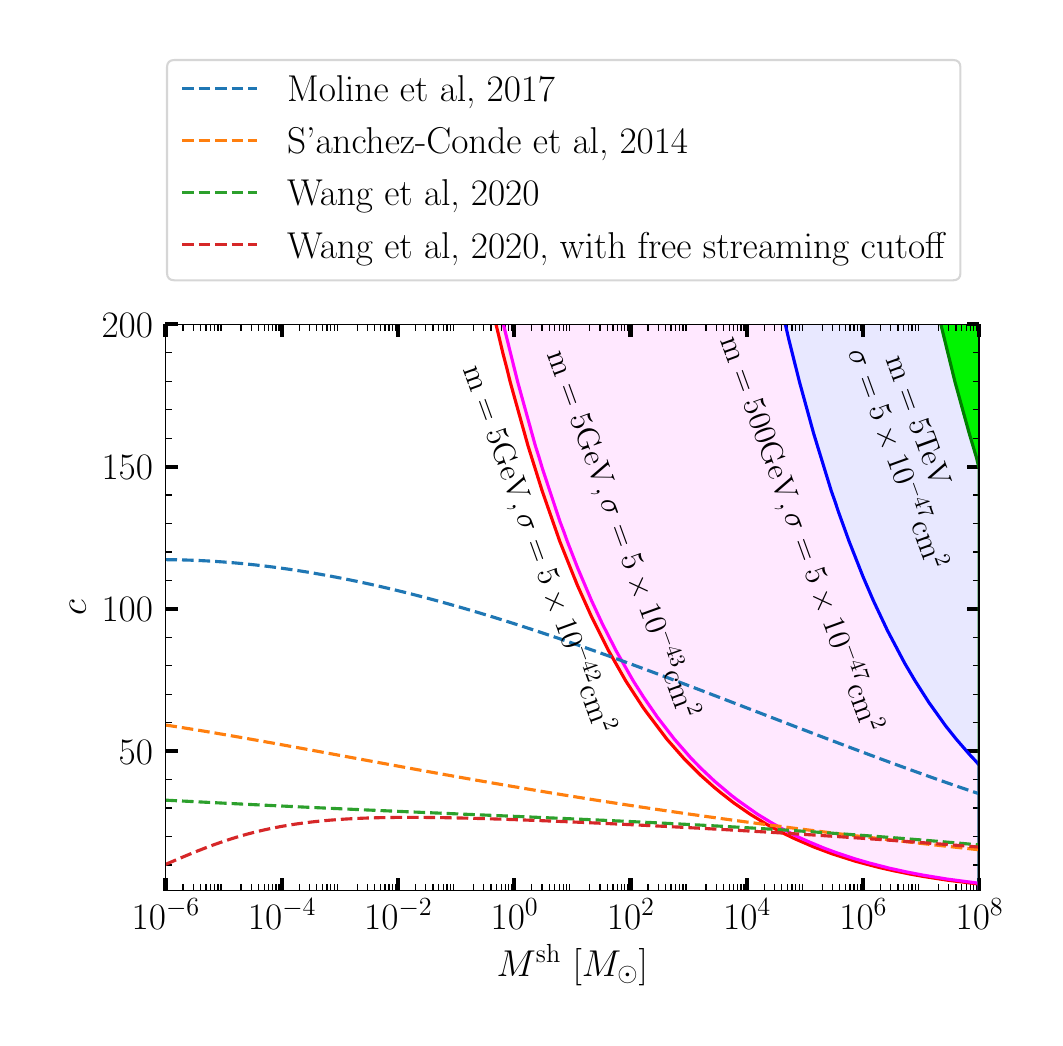}
 \caption{The potential constraints on the mass-concentration plane for specific DM models, with dashed lines representing analytical expressions from previous studies. The shaded areas, differentiated by distinct color, correspond to DM models characterized by varying masses and cross-sections. As the mass decreases and the cross-section increases, the sensitivity region covered grows larger. It appears that only lighter DM models are suitable for testing some of the existing analytical results. }
  \label{fig:CM}
 \end{center}
 \end{figure}

In Fig.~\ref{fig:CM}, we show our results as sensitivity regions on the mass-concentration plane from a subhalo encounter event. 
Analytical results from previous studies are represented with dashed lines, and we picked four different sets of masses and cross-sections for DM with corresponding sensitivity regions shaded by different colors. Three of such regions are still allowed, while $m=500~\rm{GeV}, \sigma = 5\times 10^{-47}\rm{cm}^2$ has been excluded by the latest LZ result \citep{mount2017luxzeplinlztechnicaldesign, aalbers2024darkmattersearchresults}. This excluded set of parameters is chosen in order to compare how the projected sensitivity regions change with mass or cross-section. It appears that lighter DM models with larger cross section can cover a meaningful amount of region that can test existing analytical $c(M)$ relations. 

\begin{figure}[t]
\begin{center}
  \includegraphics[width=0.95\linewidth]{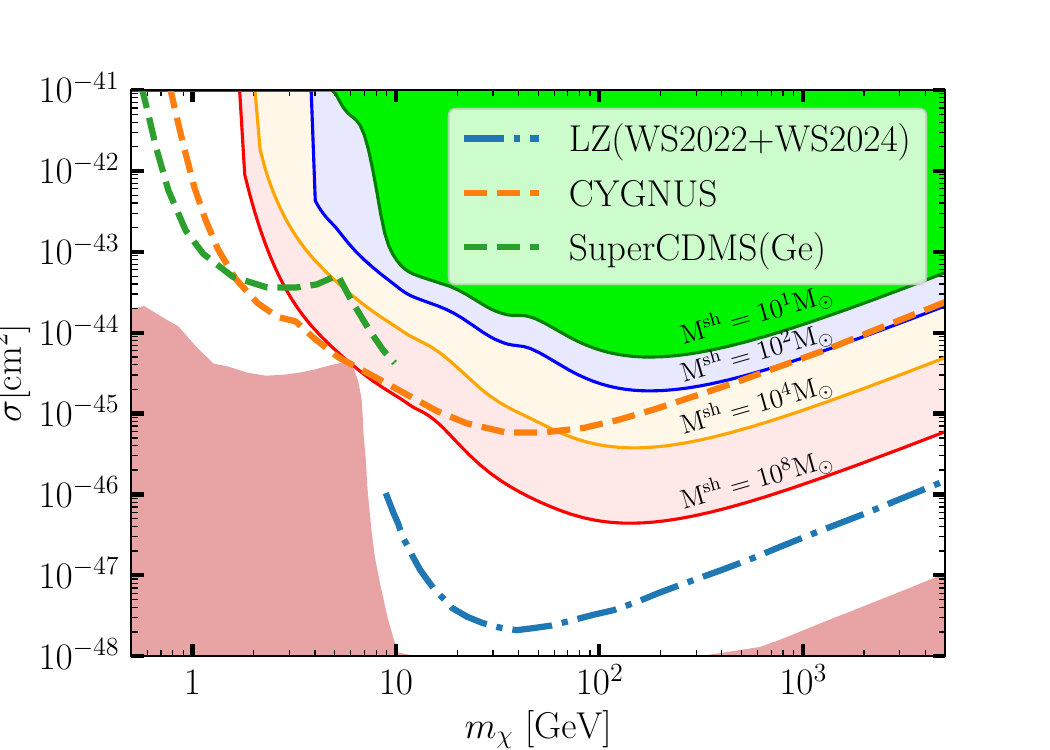}
 \caption{DM mass and cross-section needed to detect a subhalo flyby event with different subhalo mass shaded by red, yellow, blue, and green. The latest LZ result~\citep{mount2017luxzeplinlztechnicaldesign, aalbers2024darkmattersearchresults} is represented by a dashdot line that ends at 10 GeV as presented in the published result. The projected sensitivities for CYGNUS~\citep{vahsen2020cygnusfeasibilitynuclearrecoil} and SuperCDMS(Ge)~\citep{Agnese_2017} are represented by orange and green dashed lines respectively. The neutrino fog~\cite{Monroe_2007, Strigari_2009} is marked by the dark red region.}
  \label{fig:MS}
 \end{center}
 \end{figure}

To further study how variation of DM mass and cross-section affects the detectable subhalo mass, we can instead fix the mass-concentration relation, assuming the $c(M)$ proposed by Moline et al.~\cite{Molin__2017}. We can then invert Fig.~\ref{fig:CM}. The result is shown in Fig.~\ref{fig:MS}. The DM mass and cross-section needed for subhalo masses $10^8 M_{\odot}$, $10^4 M_{\odot}$, $10^2 M_{\odot}$, and $10^1 M_{\odot}$, are shaded in red, yellow, blue, and green respectively. The latest LZ result \citep{mount2017luxzeplinlztechnicaldesign, aalbers2024darkmattersearchresults} is represented by a dashdotted line. The projected sensitivities for CYGNUS~\citep{vahsen2020cygnusfeasibilitynuclearrecoil} and SuperCDMS(Ge)~\citep{Agnese_2017} are represented by orange and green dashed lines respectively. The neutrino fog~\cite{Monroe_2007, Strigari_2009} is marked by the dark red region. 

From Fig.~\ref{fig:MS}, we can see that, for the subhalo mass range of interest with the given $c(M)$, a subhalo flyby event can be detected for light DM candidates with $m_{\chi} <10$ GeV. For heavier masses, the parameter space is excluded by the LZ results \citep{mount2017luxzeplinlztechnicaldesign, aalbers2024darkmattersearchresults}. A detection of a flyby event would be more sensitive than existing constraints within the mass range 2~GeV $< m_{\chi} <10$ GeV; such range will only be explored by future experiments (e.g. CYGNUS~\citep{vahsen2020cygnusfeasibilitynuclearrecoil}, and SuperCDMS(Ge)~\citep{Agnese_2017}).  In addition, if there are subhalos with higher concentrations, for example in the cases of extremely dense objects like Axion miniclusters or dark compact objects, then detectable flyby events can potentially overcome the LZ bound, but then their mass functions would be different and require further study. For spin-dependent interactions, as discussed in ~\cite{Drukier_2019}, the results will be similar to the spin-independent ones we show here but with stronger dependence on the chemical composition of the target minerals. \cite{Drukier_2019} found that it would generally be hard to find good target materials as it would require an unpaired neutron, and such isotopes are rare in nature.

\section{\label{sec:outlook}Outlook with Alternative DM Models}

Paleo-detectors can also be applied to study physical scenarios outside of typical $\Lambda$CDM. Even though $\Lambda$CDM has been effective at describing the large-scale structure of the universe, there are a few remaining problems at small scale in $\Lambda$CDM~\cite{Governato_2010, Bullock_2017, 2019MNRAS.487.1380G, Relatores_2019, Verde_2019, Hildebrandt_2020}. A dark sector that mirrors the Standard Model could potentially alleviate these problems by introducing self-interactions between DM particles~\cite{bansal2023precisioncosmologicalconstraintsatomic} and is also well-motivated from particle physics to solve the Hierarchy problem~\cite{Bansal_2022}. 

\subsection{Atomic Dark Matter}
Atomic dark matter (ADM) is the simplest realization of such mirror dark sector model with a dark proton, a dark electron, and a dark photon~\cite{Cline_2022, bansal2023precisioncosmologicalconstraintsatomic}. In these models, it was derived~\cite{Fan_2013, Agrawal_2017, Rosenberg_2017} and then later verified in simulations that DM might form a disk-like structure at the center of the halo with an increased population of substructures within this dark disk~\cite{gemmell2023dissipativedarksubstructureconsequences, roy2024aggressivelydissipativedarkdwarfseffects}. ADM would also allow another type of substructure, called dark compact object, which is a compact DM clump with high concentration that could be thought of as a dark planet~\cite{gemmell2023dissipativedarksubstructureconsequences, roy2024aggressivelydissipativedarkdwarfseffects}. These novel substructures are a great target to look for using Paleo-detectors since they are sensitive to high concentration substructures. However, the size of the dark disk is quite sensitive to ADM parameters, and thus, whether the dark disk can extend to the solar neighborhood and boost the local number density of substructures with dark compact objects requires further studies on ADM models. 

\subsection{Strongly-Interacting Subcomponent of DM}
Another scenario that the Paleo-detectors might be exceptionally applicable for is to constrain strongly-interacting sub-component of DM~\cite{PhysRevD.41.2388, PhysRevD.41.3594, Farrar_2003, Zaharijas_2005}. A small fraction of DM, potentially down to $10^{-8}$ to $10^{-10}$, could be strongly interacting. This sub-component would be hard to constrain due to its low abundance, even though its cross-section could be tens of orders of magnitudes stronger than typical direct detection limits. 

It is proposed that these types of DM can be trapped by Earth's gravitational potential in large quantities that could be up to 15 orders of magnitude higher than the local DM density~\cite{McKeen_2022}. However, the bounded DM particles are already thermalized with only a small amount of kinetic energy around a fraction of eV. This amount of energy is too low to be detected via direct detection methods, even with the boosted density.

There have been various proposals to constrain this type of DM, including probing signals of DM upscattered in nuclear reactors, or searching for annihilation signals into neutrinos~\cite{McKeen_2022, McKeen_2023, ema2024probingearthbounddarkmatter, Pospelov_2024}. Yet, these searches are limited to probing DM with masses around $1\sim 10$ GeV. For lighter DM, the capture rate decreases, or their kinetic energies fall below the detector threshold. For heavier DM, these particles would sink towards the center of the Earth, leaving insufficient DM near the surface.

The searches for light strongly-interacting DM do not pose a challenge for Paleo-detectors, as they can record the capturing event, in which strongly-interacting DM behaves similarly to CDM, with the same velocity distribution but more scatterings. This would be in contrast to already captured DM, which would have been already thermalized, and thus has a lower velocity dispersion. We can still use Eq.~\ref{dRMW} as a first order estimate of the scattering rate. Since the event rate is proportional to $\sigma^{\rm{SI}} \rho_{\chi}$, there is a degeneracy between the cross-section and the local DM density. For larger cross-sections, we can probe an even smaller subcomponent. As a rough estimate, for a subcomponent with a fraction of $10^{-9}$, we can constrain the cross-section above around $10^{-34}~\rm{cm}^2$, which suggests a potential to cover some of the still open parameter space in the mass-concentration relation. However, Paleo-detectors might be subject to the same problem as direct detection experiments; if the cross-section becomes too large, DM might be shielded by the Earth's surface and not able to reach underground where the cosmic background is low. However, Paleo-detectors, benefiting from longer accumulation times, are more likely to record such events even at higher cross-sections. Nonetheless, detailed studies are required to identify the upper bound in this scenario.

\section{\label{sec:conclusion}Conclusions}

In this work, we estimated the yearly encounter rate of low mass subhalos for Earth and derived constraints on mass-concentration relation from a subhalo encounter event using Paleo-detectors.
Our analysis predicts $10^{-8}$ subhalo encounter events per year within a subhalo mass range of $10^{-6}M_{\odot}$ to $10^5M_{\odot}$. Unlike traditional direct detection experiments, we showed that, due to their extended integration time of Paleo-detectors, it is possible that subhalo encounter events were recorded by Paleo-detectors. These signatures can be used to constrain subhalo properties, like the mass-concentration relation. We found that low mass DM of 5 GeV and cross sections of $10^{-42}$cm$^2$ can be used to probe subhalos with masses ranging from $\sim 10^5 M_{\odot}$ down to $100~M_{\odot}$, depending on the assumed mass-concentration relation. Fixing such relation, we find that this technique might even be a more sensitive probe to light DM models (with masses $\mathcal{O}$(1 GeV)) than the current direct detection results in that mass range.

There are a few assumptions considered in this work that can be improved upon. First, the effect of tidal disruption due to the presence of the disk of the MW is assumed to cause a reduction of the surviving subhalos by a factor of three, although detailed studies are required to obtain a proper radial dependence of such reduction. Additionally, tidal effects get stronger towards the center of the galaxy, and thus subhalos that are still present near the center of the MW are generally more massive and have higher concentration. Therefore, the mass function will depend on the distance to the center of the MW, and Eq.~\ref{eq:subhalo_mass_function} and Eq.~\ref{dn} will require some modification. Secondly, subhalos were more numerous at higher redshifts. For example, in \cite{Barry_2023}, it is shown that there would be about twice more subhalos at $z=0.5$ which is comparable to the age of the Earth. This will slightly increase the chance of a subhalo encounter event by up to a factor of two. Finally, we also assumed a conventional mass-concentration relation to show that a detection of a flyby event is probable for low mass DM models. If there exist substructures with high concentrations, then such a limit can be improved, and we might be able to detect such events with heavier DM mass even below the current LZ bound. We expect that the above modifications will only enhance the reach of Paleo-detectors.  Ref.~\cite{fung2025refiningsensitivitynewphysics} states, however, that the number of tracks in other previous works could be overestimated. Further studies are required to explore how the combined effects will modify the results. 

Although the analysis of this work holds true for $\Lambda$CDM models, the parameter space will be altered for different DM models, such as atomic DM or or a strongly-interacting subcomponent of DM, and detailed analyses should therefore be conducted in these scenarios. In ADM models, we expect the presence of dark clumps, which in some cases might form a dark disk (the size of which depends on the details of the model). Such an excess of dark dense structures, which might escape the lifetime of a standard direct detection experiment of $\mathcal{O}(10)$ years, can be probed using the extended integration time of Paleo-detectors. In the strongly-interacting DM, Paleo-detectors can record the subcomponent of DM as it gets trapped by the Earth's gravitational field, with more scatterings compared to CDM, over extended periods of time. Such alternative DM model studies, as well as others, deserve detailed analyses, which will be tackled in future work.

In this work, we demonstrated that Paleo-detectors are an important novel technique that can be complementary to traditional direct detection experiments. It can not only help constrain DM parameter space, but also open up new avenues to study halo substructures through geological history, with potential applications to other scenarios outside of $\Lambda$CDM. Paleo-detectors should therefore be considered a new piece of our arsenal as we search for the particle nature of DM, providing valuable insights about the subhalo mass function, and more generally, the accretion history of the Milky Way.

\section*{Acknowledgments}

We would like to thank  He Feng, Marianne Moore, Sandip Roy, and Tracy Slayter for helpful conversations.
XZ is partially supported by DOE award DE-SC0024112.
LN is supported by the Sloan Fellowship, the NSF CAREER award 2337864, NSF award 2307788, and by
the NSF award PHY2019786 (The NSF AI Institute
for Artificial Intelligence and Fundamental Interactions,
\url{http://iaifi.org/}). DE acknowledges the support of the Australian Research Council through project number DP220102254.

\appendix

\section{Note on Mass Scaling of the Differential Encounter Rate}
\label{app:scaling}

In this Appendix, we present how the mass scaling of the differential encounter rate is analyzed. Looking at Eq.~\ref{eq6} and Eq.~\ref{dn}, we can find the scaling of the differential rate in terms of the subhalo mass as

\begin{equation}
   \frac{dR_{\sub}}{dM_{\sub}}\sim M_{\sub}^{-1.9}\times b_{\rmmax}^2(M_{\sub}).  
\end{equation}

From Eq.~\ref{bmax}, we know that $b_{\rmmax} \sim \sqrt{r_c^2+\frac{2GM_E}{v_0^2}r_c}$. So $b_{\rmmax}$ scales as $r_c$ for large $r_c$ and $\sqrt{r_c}$ for small $r_c$. Since $r_c \sim M_{\sub}^{1/3}$, then towards the lower mass end

\begin{equation}
   \frac{dR_{\sub}}{dM_{\sub}}\sim M_{\sub}^{-1.9}\times b_{\rmmax}^2(M_{\sub})\sim M_{\sub}^{-1.9}\times r_c \sim M_{\sub}^{-1.6},  
\end{equation}

and towards the higher mass end, 
\begin{equation}
   \frac{dR_{\sub}}{dM_{\sub}}\sim M_{\sub}^{-1.9}\times b_{\rmmax}^2(M_{\sub})\sim M_{\sub}^{-1.9}\times r_c^2 \sim M_{\sub}^{-1.3}.  
\end{equation}

The change of slope happens around $r_c \sim \frac{2GM_E}{v_0^2}$, which is around $6.5 \times 10^{-13}$~pc for $v_0 \sim 200\rm{km/s}$ and far smaller than the smallest radius we considered. So we are always in the high mass regime in this study. 

\bibliography{apssamp}

\end{document}